# Multi-modal clothing recommendation model based on large model and VAE enhancement

BINGJIE HUANG∗

Independent researcher, Sunnyvale, CA, USA

bingjiehuang1998@gmail.com

QINGYI LU

Brown University, Department of Computer Science, Providence, RI, USA

lunalu9739@gmail.com

SHUAISHUAI HUANG

Independent researcher, Hefei, Anhui, China

scu.hss@gmail.com

XUE-SHE WANG

Duke University, Pratt School of Engineering, Durham, NC, USA

wxs.research@gmail.com

HAOWEI YANG

University of Houston, Cullen College of Engineering, Houston, Texas, USA

hyang38@cougarnet.uh.edu

Accurately recommending products has long been a subject requiring in-depth research. This study proposes a multimodal paradigm for clothing recommendations. Specifically, it designs a multimodal analysis method that integrates clothing description texts and images, utilizing a pre-trained large language model to deeply explore the hidden meanings of users and products. Additionally, a variational encoder is employed to learn the relationship between user information and products to address the cold start problem in recommendation systems. This study also validates the significant performance advantages of this method over various recommendation system methods through extensive ablation experiments, providing crucial practical guidance for the comprehensive optimization of recommendation systems.

# 1 INTRODUCTION

The fashion industry's rapid development has resulted in an overwhelming amount of data and information for users when selecting clothing. The abundance of clothing product information makes it difficult for consumers to choose among different clothing options when shopping online. In the era of information explosion, accurately retrieving clothing products that meet users' needs has become an urgent research problem to be solved. Therefore, it is necessary to adopt an automatic and effective method that can enable consumers to purchase suitable clothes more efficiently. The intelligent clothing recommendation system can analyze a series of user characteristics and rank the most suitable products according to certain rules, giving more importance or attention to clothing products that users may be interested in or most suitable for, thereby achieving effective information filtering to achieve the effect of recommendation [1].

Typical recommender systems use two types of data: rating or purchase behavior reflecting attribution information of products and users, and keywords or textual information that record user-item interactions. Generally, recommender systems rely on various types of input. Explicit feedback data is usually structured and can be directly modeled without special processing. This data primarily records key textual information and details of user-item interactions. For instance, a common example is users expressing their preferences for products through likes or star ratings [2]. In the absence of explicit feedback, recommender systems often infer user information from implicit feedback, such as purchase history or watchlists for clothing items. Most recommender systems can be categorized into three types: content-based recommender systems, collaborative filtering recommender systems, and hybrid recommender systems. One of the earliest and most widely used techniques in recommender systems is collaborative filtering. It estimates and generates recommendations based on the commonalities and similarities of users with similar purchase behaviors and information [3]. In addition to collaborative filtering models, content-based filtering is another significant type of recommender system. Content-based recommender systems recommend other items similar to a user's preferences by considering the contributions of a particular item, focusing mainly on a single user's ratings rather than multiple users' ratings. Moreover, content-based recommender systems rely more heavily on the higher-level features of users and items for predictions compared to collaborative filtering [4].

Both collaborative filtering and content-based recommender systems have their limitations. For instance, collaborative filtering systems tend to focus more on structured data, while content-based recommender systems may overlook the similarity of preferences among individuals, impacting the accuracy of predictions [5]. Hybrid recommender systems integrate the advantages of both types of recommender systems to improve the comprehensiveness of algorithms. However, these systems face issues such as cold start, sparsity, and outdated information, which limit their predictive performance [6]. The problem solved in this paper is as follow:

Q1:how to set up a multimodal model to achieve good performance during cold start (lack of data)?

This study develops an integrated recommendation model combining a large-scale model and variational autoencoder to address performance degradation issues in cold start or low data volume scenarios. The contributions of this model are threefold:

1. LLMs is introduced to encode and fuse different modalities to improve the accuracy of information expression

2. Neural collaborative filtering is used to replace traditional filtering methods to improve information interaction

3. VAE is used to obtain pseudo-label enhanced data to alleviate the cold start problem

This method effectively addresses the current cold start and accuracy issues in recommendation systems, significantly contributing to the improvement of recommendation systems, and introduces a new paradigm that integrates multiple modalities to enhance recommendation accuracy.

## 2 LITERATURE REVIEW

### 2.1 Application of Deep Learning in Recommendation

With the emergence of deep networks, the application of artificial intelligence in education [7], medical care [8], business [9], credit modeling [10], chip design [11] and other fields has made great progress. In traditional collaborative filtering algorithms, matrix factorization commonly uses the inner product to represent interactions. In theory, neural networks can fit any function, thus allowing them to learn from data to replace the inner product for modeling latent features of users and items. This approach is referred to as Neural Collaborative Filtering [12]. In addition to the improvement of the matrix decomposition algorithm, in order to overcome the above-mentioned obstacles of traditional models and achieve more accurate recommendations, researchers are considering the possibility of using deep learning for recommender systems. With deep learning, hidden information can be obtained from including contextual, textual, and visual inputs have detected nonlinear relationships between users and recommended items [13]. For example, Bhasker et al. proposed a novel deep learning hybrid recommendation algorithm that uses specific embeddings and deep neural networks to characterize users and items in order to learn nonlinear latent factors [14] Lin et al. developed a clothing recommendation system based on a neural network model, which firstly uses a convolutional neural network for gender recognition, and then modules such as Dlib and GoogleNet for clothing attribute recognition [15]. With the rise of training large models, the accuracy of this recommendation system has been further improved, such as directly recommending products through large models [16] or training large models to provide embedded vectors for prediction [17]. However, although the semantics of large models greatly enhance the effectiveness of recommendation systems, there are still challenges in cold start and fusion methods [18]. Large models usually require a lot of data to generate effective vector representations, but it may be difficult to have enough data to support subsequent training and fine-tuning. At the same time, large models are usually difficult to update frequently, and retraining or fine-tuning a large model also requires a lot of resources. This may reduce the timeliness of the recommendation system in a rapidly changing recommendation environment.

### 2.2 Multi-modal Fusion and cold start

Modal fusion methods typically include early fusion and late fusion. Early fusion refers to the merging of data from multiple modalities prior to model input, allowing the model to process interactions between different modalities from the outset. Late fusion, on the other hand, involves the combination of independent judgments from different modalities at the final decision stage of the model [19]. While early fusion is simple and efficient, it can lead to incomplete utilization of semantic information due to the compression of features with different scales at the initial stage of fusion. In contrast, late fusion allows for the independent training of each modality, applying the most appropriate techniques to handle different types of modalities. However, integrating outputs after separate training can present challenges in capturing the deep interactions between different modalities. Additionally, there exists an intermediate fusion approach, which serves as a compromise by combining multimodal features at certain intermediate layers to capitalize on the advantages of both early and late fusion.

Techniques employed in intermediate fusion include concatenation, weighted summation, and neural networks.

The cold-start problem in multimodal fusion generally refers to the initial stage of the model, due to a lack of sufficient multimodal data or user interaction history, the model struggles to effectively utilize information from each modality for accurate predictions or recommendations [20]. When This issue may arise from various challenges, including data sparsity, modality inconsistency, and insufficient model training [21]. To address the cold-start problem in recommendation systems, some studies have employed enhanced collaborative filtering [22], transfer learning [23], and pseudo-label generation techniques [24]. We adopt a pseudo-labeling method for enhancement.

## 3 MODELING

In general, our overall model meets the requirements of cross-modal tasks and neural collaborative filtering algorithms. Our model process is shown in Figure 1, which mainly includes data organization, cross-modal embedding representation, model training and evaluation. We collect text and image information of users and products and send them to the cross-modal communication block. The data generated in the communication block will then be sent to the neural collaborative network for comparative prediction.

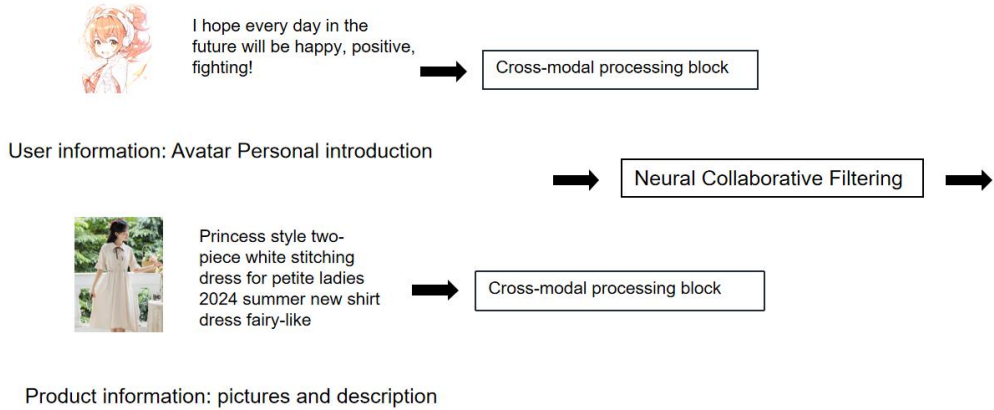

Figure 1. Multimodal Recommendation System Model Process

The structure of the cross-modal communication block is shown in Figure 2. The cross-modal communication block includes an output layer, a pre-trained model processing module, a fusion module, a variational encoder structure and an output layer. The text and image data will be sent to the LLMs for embedding vector encoding, then output by the fusion layer, and finally output to the variational encoder to improve the generalization of the data before the final output.

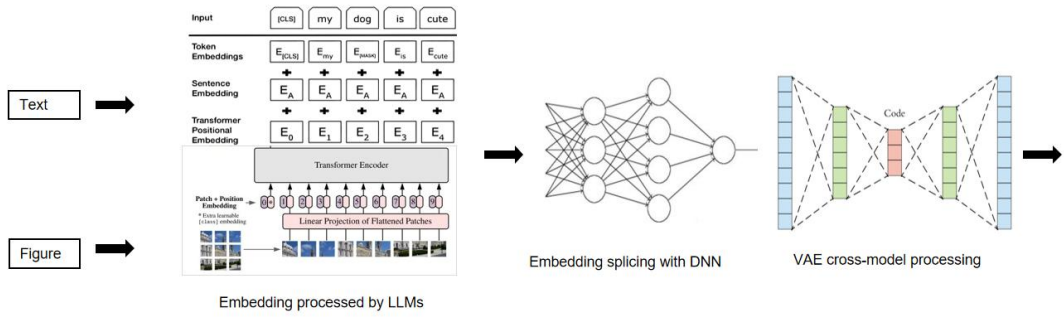

Figure 2. Cross-modal processing module

The introduction of BERT marked a significant advancement in natural language processing [25], as it employs a bidirectional encoder architecture that allows the model to consider both preceding and succeeding words in a sentence simultaneously. This contrasts with traditional models that process text in a single direction, and it has been widely demonstrated that BERT effectively captures contextual and semantic relationships in text, thereby providing a more comprehensive understanding of context. The embedding components of BERT include word embeddings, segment embeddings, and position embeddings. In essence, word embeddings map each word individually into a vector within a high-dimensional space. The segment embeddings allow BERT to differentiate and process single texts or pairs of texts, thereby enabling the understanding of semantic information at the sentence level. The position embeddings provide sequential information to the structure, allowing the model to mark the position of words in a sentence, which aids in further processing at a higher level. Finally, the CLS token at the beginning of the input sequence represents the final hidden state in the embedding vector, which is commonly used as the representation of the entire input sequence.

The training of the BERT model can be divided into two distinct phases. The Masked Language Model (MLM) and the Next Sentence Prediction (NSP) are two key components of the BERT training process. The MLM mechanism randomly selects a subset of words from the input text and replaces them with a special mask token. Subsequently, the model makes a prediction regarding the masked words, utilizing contextual information to facilitate the acquisition of more nuanced language representations. In contrast, the NSP approach involves the selection of two consecutive sentences from the training corpus as positive samples, with the process then repeated to identify negative samples. Following the labeling of the sentences, the model is trained to output a binary classification result, indicating whether the second sentence follows the first sentence. Vision Transformer (ViT) is another method that applies the Transformer model to the image domain. It operates on principles similar to BERT. The main structure of ViT is a multi-layer Transformer encoder, which first divides the input image into fixed-size patches, converts these pixel blocks into one-dimensional sequences, and adds word embeddings, segment embeddings, and position embeddings to form new feature vectors. It then utilizes the self-attention mechanism to calculate the relevance within the input sequence and applies attention to different parts of the input sequence to capture hidden key information in the image. These input sequences are processed through multiple Transformer encoder layers, each containing multi-head self-attention mechanisms and MLPs [26]. Variational Autoencoder (VAE) is a generative model that combines the structure of an autoencoder with variational Bayesian inference which shown as Figure 3 [27]. It learns a representation of the latent distribution of the input data through an encoder and then generates new samples from the latent space using a decoder. Specifically, it maps input data to a latent space that approximates a Gaussian distribution, samples latent variables from this distribution, and reconstructs the data from these sampled latent variables.

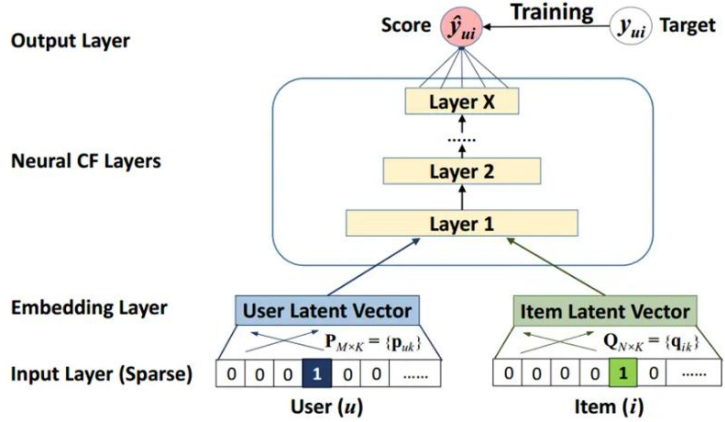

Figure 3. Structure of VAE

The data processed by the variational autoencoder, which includes both user and item data, is then fed into a neural matrix factorization module. This module combines Generalized Matrix Factorization (GMF) with neuron units based on MLPs. The GMF captures linear interactions using the dot product of traditional user and item embedding vectors. The user and item embedding vectors are also input into a multilayer perceptron to capture additional nonlinear interactions simultaneously. Finally, the output results of the processed user and item embedding vectors are combined through an MLP in the fusion layer. In NCF, $\widehat{y_{ui}}$ represents the predicted rating of user $u$ for item $i$.. $h$, $W$, and $b$ are NCF model parameters. $\sigma$ and $t$ represents an activation function. ∘ is the dot product. $p_u$ and $q_i$ stand for the embedding vectors for user $u$ and item $i$.

$$\widehat{y_{ui}} = \sigma\big(h^T t(p_u \circ q_i) + W[p_{uq_i}] + b\big) \#(1)$$

## 4 EVALUATES MATREIX

The paper evaluates the performance of models using three common methods: Mean Squared Error (MSE), Average Accuracy, and Precision@K. MSE, represented in Formula 2, assesses the regression performance by calculating the average squared difference between the predicted and actual values. $n$ is the total number of sample points, $\hat{r}_i$ is the predicted result of the i-th observation predicted by the model, and $r_i$ is the actual value of the i-th observation.

$$\text{MSE} = \frac{1}{n}\sum_{i=1}^{n}(\hat{r}_i - r_i)^2 \#(2)$$

Precision@K evaluates recommendation systems by measuring the ratio of relevant items in the top K recommendations to the total number of recommendations, as shown in Formula 3. $U$ represents users, $R_u$ is standing for the interaction between the user and the item, and $R_u \cap \widehat{R_{u,K}}$ is the results of recommended items the user put the relatively high ranking.

$$\text{Precision@K} = \frac{1}{|U|} \sum_{u \in U} \frac{|R_u \cap \widehat{R_{u,K}}|}{K} \#(3)$$

Normalized Discounted Cumulative Gain (NDCG) is a common metric for assessing the relevance and ranking of results in recommendation systems. DCG@K measures the cumulative gain of the top K recommended items for a user. while IDCG@K represents the maximum possible value of DCG@K under ideal conditions. This weighting method ensures that higher-ranked correlations receive greater importance, which affects the overall ranking significance. CG computes the total correlations in the recommendations without considering their positions, as illustrated in Formula 4, while DCG accounts for the position loss factor in the recommendation results.

$$\text{NDCG@K} = \frac{1}{|U|} \sum_{u \in U} \frac{\text{DCG@K}}{\text{IDCG@K}} \#(4)$$

## 5 EVALUATES MATREIX

In order to adapt our proposed model, we refer to the Fashion-mnist dataset [28] and obtain data from Taobao, China's largest online shopping website. We created a multimodal dataset that includes user avatars and introductions, clothing pictures and introductions. The introduction has been screenshotted to ensure that all text information in the picture only contains pure picture information. We still prepared a more detailed feature for subsequent ablation experiments, which includes the number of clicks on the product page by the user, the age and gender of the user's Taobao account opening, the user's comments on the product, clothing category, color, style and other features. This dataset has 11 features and a total of 82 data.

In all experiments, all MLP models were configured as three-layer structures, with the first layer comprising 128 neurons and the second layer comprising 64 neurons. All neural networks were trained for 10 epochs, the number of recommended items in the recommendation system $K$ was set to 5, and cross-validation was employed for training. To assess the efficacy of the proposed model, we conducted a series of ablation experiments and compared the results with those of the following models. We use the following mainstream models for comparison:

1. Matrix factorization model (MF) [29]
2. Neural matrix factorization model (NMF) [30]
3. multimodal recommendation mode CNN(CNN) [31]
4. multimodal recommendation model of DNN (DNN) [32]
5. CLIP [33]
6. VilBERT [34]

MF and NMF will directly use MLP to process text and image data. Image data is processed using Scale-Invariant Feature Transform, a non-deep learning method, while text is processed using Term Frequency-Inverse Document Frequency to obtain vector embedding. It should be noted that all the benchmark models only involve modal fusion and only contain text and image content. The fusion of non-structural content will be performed separately in the ablation experiment. Non-structural content will be additionally entered into the splicing after multi-modal merging and merged through MLP.

# 6 EXPERIMENTS

Table 1: Experiments result

| Models | MSE | Precision@K | NDCG |
|---|---|---|---|
| Muti (ours) | 0.44 | 0.92 | 0.93 |
| MF | 1.55 | 0.75 | 0.71 |
| NMF | 2.75 | 0.85 | 0.67 |
| CNN | 0.56 | 0.93 | 0.85 |
| DNN | 0.51 | 0.92 | 0.79 |
| CLIP | 0.37 | 0.91 | 0.93 |
| ViLBERT | 0.32 | 0.90 | 0.90 |
| Muti-with structure data | 0.43 | 0.91 | 0.90 |
| Muti-with no VAE | 0.44 | 0.88 | 0.90 |

Table 1 presents the performance of various models across three metrics: MSE, Precision@K, and NDCG. First, in terms of MSE, ViLBERT performs the best with the lowest MSE of 0.32, followed by CLIP with an MSE of 0.37. The MSE of Muti (ours) and its variants ranges from 0.43 to 0.44, indicating satisfactory performance. In contrast, traditional MF and NMF models exhibit significantly higher MSE values of 1.55 and 2.75, respectively. Secondly, regarding Precision@K, CNN and CLIP excel with the highest Precision@K values of 0.93, followed closely by Muti (ours) and its variants, which achieve Precision@K values ranging from 0.88 to 0.92. The MF model, however, has the lowest Precision@K of only 0.75. Lastly, in terms of NDCG, CLIP and Muti (ours) lead with the highest NDCG values of 0.93, with ViLBERT close behind at 0.90. Muti (ours) and its variants exhibit consistent NDCG values between 0.90 and 0.93, demonstrating balanced overall performance. On the other hand, the NMF model has a relatively low NDCG value of 0.67, indicating poorer performance.

The ablation results shown in Table 2 provide a nuanced understanding of how different fusion strategies, as well as the inclusion or exclusion of the variational autoencoder (VAE), affect the performance of the Muti model on multiple evaluation metrics. We are able to find that it is first the intermediate fusion strategy combined with the VAE that most effectively captures and integrates multimodal interactions, resulting in accurate predictions and high-quality rankings. This configuration appears to strike a balance between early and late fusion, allowing the model to benefit from early feature learning while also utilizing late integration to improve ranking. In addition, the Muti-with no VAE-Early Fusion and late Fusion variants with no VAE and no late Fusion performed at 0.51 and 0.58, respectively, with lower Precision@K and NDCG values, further demonstrating the importance of VAE and fusion strategies. The reduced performance of these variants reinforces the conclusion that both late Fusion and the absence of VAE lead to suboptimal modal integration, which in turn leads to poorer prediction and ranking results.

Table 2: Ablation Experiments result

| Models | MSE | Precision@K | NDCG |
|---|---|---|---|
| Muti (with VAE -middle Fusion) | 0.44 | 0.92 | 0.93 |
| Muti-Early Fusion | 0.57 | 0.62 | 0.65 |
| Muti-late Fusion | 0.51 | 0.85 | 0.88 |
| Muti-with structure data | 0.43 | 0.91 | 0.90 |
| Muti-with no VAE- | 0.44 | 0.88 | 0.90 |
| Muti-with no VAE- Early Fusion | 0.51 | 0.82 | 0.86 |
| Muti-with no late | 0.58 | 0.82 | 0.85 |

## 7 DISCUSSION

In total, VilBERT and CLIP demonstrate outstanding performance across all metrics, especially in MSE and NDCG. Muti (ours) and its variants show balanced and commendable performance across all evaluated metrics, making them reliable choices. In contrast, traditional MF and NMF models are relatively less competitive in modern applications. In addition, we conducted another round of ablation experiments to verify whether mid-term fusion is the best. In our ablation experiment, we added early fusion and late fusion. We directly used simple concatenation for early fusion. Then we put the VAE module at the output and added a fully connected network. Finally, we experimented with late fusion.

Through a series of experiments, we have reached three preliminary insights: First, multimodal models based on large language models perform significantly better than traditional methods and deep neural network-based methods. This advantage is primarily attributed to the richness of the vector embeddings. Large language models can generate embeddings with richer semantic information, thereby performing better in multimodal tasks. Second, although our model also uses large language models for vector embeddings, its performance is slightly inferior to mainstream models like CLIP and Vilbert. This is likely because our model lacks the crucial step of explicit feature alignment. Although we performed image-text alignment through MLPs, the absence of an explicit self-supervised learning process may have led to performance loss.Third, fortunately, our ablation experiments demonstrate that the introduction of the VAE structure can significantly reduce the dependency on data.

Experiments show that incorporating VAE significantly enhances the model's predictive capabilities. However, we remain uncertain about the profound impact of VAE on the model, as VAE can distort images during generation, such as sharpening colors. These data augmentation behaviors may not necessarily provide guidance in real-world recommendation systems. In addition, the extent to which the VAE model is added to the model should also be further considered, which is related to the overall consideration of the entire multimodal construction [35]. Additionally, we conducted ablation experiments to assess the impact of unstructured data on the model. We integrated unstructured data into the VAE-generated embeddings and performed dimensionality reduction to revert to the original dimensions. Experiments show that adding extra information does positively affect the model's performance, but the effect is still inferior to models like CLIP that only use explicit alignment methods. This further confirms the importance of explicit alignment in multimodal recommendation systems. However, as with VAE, how this explicit modality alignment should be reasonably built into the model remains an issue that needs further discussion [36].

In the field of data augmentation, a commonly used approach is to establish relationships using graph-based [37] methods or to map with Apriori rule algorithms [38]. These methods rely on predefined rules, making the data augmentation process interpretable and the results highly predictable. However, unlike rule-based cold-start augmentation, data augmentation through VAE is a data-driven approach, offering greater flexibility and unpredictability. It is particularly important to note that for multimodal models, which inherently possess higher complexity, rule-based approaches are highly likely to fail during the multimodal fusion process. VAE, therefore, provides a solution for cold start in complex multimodal fusion in recommendation systems. In future experiments, further in-depth research will be conducted to explore the differences in effectiveness between the two methods.

## 8 CONCLUSION

This paper proposes an integrated recommendation model that combines large-scale models and variational autoencoders to solve the problem of performance degradation of recommendation systems in cold start or scenarios with small amounts of data. Experimental results show that multi-modal models based on large language models are better than traditional methods and deep neural network methods in terms of effect, benefiting from the richness of vector embeddings. Although the performance is slightly inferior to mainstream models such as CLIP and Vilbert, the introduction of VAE significantly improves prediction capabilities and reduces data dependence. Ablation experiments further confirm the importance of explicit alignment in multi-modal recommender systems.

## REFERENCES

[1] Aggarwal, C. C. (2016). Recommender systems (Vol. 1). Cham: Springer International Publishing.
[2] X. Su and T. M. Khoshgoftaar. 2009. A survey of collaborative filtering techniques. Advances in Artificial Intelligence, vol. 2009.
[3] J. L. Herlocker, J. A. Konstan, and J. Riedl. 2000. Explaining collaborative filtering recommendations. In Proceedings of the 2000 ACM conference on Computer supported cooperative work (pp. 241-250).
[4] L. Si and R. Jin. 2003. Flexible mixture model for collaborative filtering. In Proceedings of the 20th International Conference on Machine Learning (ICML-03), pp. 704-711.
[5] B. Sarwar, G. Karypis, J. Konstan, and J. Riedl. 2001. Item-based collaborative filtering recommendation algorithms. In Proceedings of the 10th International Conference on World Wide Web, pp. 285-295.
[6] J. Liu, W. H. Choi, and J. Liu. 2021. [Retracted] Personalized Movie Recommendation Method Based on Deep Learning. Mathematical Problems in Engineering, 2021(1), 6694237.
[7] Y. Luo and Z. Wang. 2024. Feature Mining Algorithm for Student Academic Prediction Based on Interpretable Deep Neural Network. In Proceedings of the 2024 12th International Conference on Information and Education Technology (ICIET) (pp. 1-5). IEEE.
[8] W. Dai, Y. Jiang, C. Mou, and C. Zhang. 2023. An Integrative Paradigm for Enhanced Stroke Prediction: Synergizing XGBoost and xDeepFM Algorithms. In Proceedings of the 2023 6th International Conference on Big Data Technologies (pp. 28-32).
[9] Y. Luo, R. Zhang, F. Wang, and T. Wei. 2023. Customer Segment Classification Prediction in the Australian Retail Based on Machine Learning Algorithms. In Proceedings of the 2023 4th International Conference on Machine Learning and Computer Application (pp. 498-503).
[10] S. Li, X. Dong, D. Ma, B. Dang, H. Zang, and Y. Gong. 2024. Utilizing the lightgbm algorithm for operator user credit assessment research. arXiv preprint arXiv:2403.14483.
[11] H. Qu, D. Ma, Z. Qi, and N. Zhu. 2024. Advanced deep-learning-based chip design enabling algorithmic and hardware architecture convergence. In Proceedings of the Third International Conference on Algorithms, Microchips, and Network Applications (AMNA 2024) (Vol. 13171, pp. 249-255). SPIE.
[12] S. Rendle, W. Krichene, L. Zhang, and J. Anderson. 2020. Neural collaborative filtering vs. matrix factorization revisited. In Proceedings of the 14th ACM Conference on Recommender Systems (pp. 240-248).
[13] R. Mu. 2018. A survey of recommender systems based on deep learning. IEEE Access, 6, 69009-69022.
[14] B. Bhasker. 2020. DNNRec: A novel deep learning based hybrid recommender system. Expert Systems with Applications, 144.
[15] Y. R. Lin, W. H. Su, C. H. Lin, B. F. Wu, C. H. Lin, H. Y. Yang, and M. Y. Chen. 2019. Clothing recommendation system based on visual information analytics. In Proceedings of the 2019 International Automatic Control Conference (CACS) (pp. 1-6). IEEE.
[16] Y. Luo, Z. Ye, and R. Lyu. 2024. Detecting student depression on Weibo based on various multimodal fusion methods. In Proceedings of the Fourth International Conference on Signal Processing and Machine Learning (CONF-SPML 2024) (Vol. 13077, pp. 202-207). SPIE.

[17] A. Xiang, Z. Qi, H. Wang, Q. Yang, and D. Ma. 2024. A Multimodal Fusion Network For Student Emotion Recognition Based on Transformer and Tensor Product. arXiv preprint arXiv:2403.08511.

[18] P. Nagarnaik and A. Thomas. 2015. Survey on recommendation system methods. In Proceedings of the 2015 2nd International Conference on Electronics and Communication Systems (ICECS) (pp. 1603-1608). IEEE.

[19] S. Li and Y. Xiao. 2024. A Depression Detection Method Based on Multi-Modal Feature Fusion Using Cross-Attention. arXiv preprint arXiv:2407.12825.

[20] C. Zhang, G. Long, T. Zhou, Z. Zhang, P. Yan, and B. Yang. 2024. When Federated Recommendation Meets Cold-Start Problem: Separating Item Attributes and User Interactions. In Proceedings of the ACM on Web Conference 2024 (pp. 3632-3642).

[21] R. Tang, C. Yang, and Y. Wang. 2023. A Cross-Domain Multimodal Supervised Latent Topic Model for Item Tagging and Cold-Start Recommendation. IEEE MultiMedia, 30(3), 48-62.

[22] I. Hossain, S. Puppala, M. J. Alam, and S. Talukder. 2024. SocialRec: User Activity Based Post Weighted Dynamic Personalized Post Recommendation System in Social Media. arXiv preprint arXiv:2407.09747.

[23] T. Iqbal, M. Masud, B. Amin, C. Feely, M. Faherty, T. Jones, et al. 2024. Towards integration of artificial intelligence into medical devices as a real-time recommender system for personalized healthcare: State-of-the-art and future prospects. Health Sciences Review, 100150.

[24] J. Bobadilla and A. Gutiérrez. 2024. Wasserstein GAN-based architecture to generate collaborative filtering synthetic datasets. Applied Intelligence, 54(3), 2472-2490.

[25] J. Devlin, M. W. Chang, K. Lee, and K. Toutanova. 2018. BERT: Pre-training of deep bidirectional transformers for language understanding. arXiv preprint arXiv:1810.04805.

[26] D. Ma, M. Wang, A. Xiang, Z. Qi, and Q. Yang. 2024. Transformer-Based Classification Outcome Prediction for Multimodal Stroke Treatment. arXiv preprint arXiv:2404.12634.

[27] M. J. Kusner, B. Paige, and J. M. Hernández-Lobato. 2017. Grammar variational autoencoder. In Proceedings of the International Conference on Machine Learning (pp. 1945-1954). PMLR.

[28] H. Xiao, K. Rasul, and R. Vollgraf. 2017. Fashion-MNIST: A novel image dataset for benchmarking machine learning algorithms. arXiv preprint arXiv:1708.07747.

[29] D. Lee and H. S. Seung. 2000. Algorithms for non-negative matrix factorization. Advances in Neural Information Processing Systems, 13.

[30] H. J. Xue, X. Dai, J. Zhang, S. Huang, and J. Chen. 2017. Deep matrix factorization models for recommender systems. In Proceedings of IJCAI (Vol. 17, pp. 3203-3209).

[31] S. S. Choudhury, S. N. Mohanty, and A. K. Jagadev. 2021. Multimodal trust-based recommender system with machine learning approaches for movie recommendation. International Journal of Information Technology, 13, 475-482.

[32] Y. Mu and Y. Wu. 2023. Multimodal movie recommendation system using deep learning. Mathematics, 11(4), 895.

[33] A. Radford, J. W. Kim, C. Hallacy, A. Ramesh, G. Goh, S. Agarwal, ... and I. Sutskever. 2021. Learning transferable visual models from natural language supervision. In Proceedings of the International Conference on Machine Learning (pp. 8748-8763). PMLR.

[34] J. Lu, D. Batra, D. Parikh, and S. Lee. 2019. ViLBERT: Pretraining task-agnostic visiolinguistic representations for vision-and-language tasks. Advances in Neural Information Processing Systems, 32.

[35] G. Chen, M. Liu, Y. Zhang, Z. Wang, S. M. Hsiang, and C. He. 2023. Using images to detect, plan, analyze, and coordinate a smart contract in construction. Journal of Management in Engineering, 39(2), 04023002.

[36] C. He, M. Liu, Z. Wang, G. Chen, Y. Zhang, and S. M. Hsiang. 2022. Facilitating smart contract in project scheduling under uncertainty—A Choquet integral approach. In Proceedings of the Construction Research Congress 2022 (pp. 930-939).

[37] C. Maier and D. Simovici. 2022. Bipartite graphs and recommendation systems. Journal of Advances in Information Technology (in print).

[38] E. Surya Negara, et al. 2023. Recommendation system with content-based filtering in NFT marketplace. Journal of Advances in Information Technology, 14(3), 518-522.